\documentclass[preprint,showpacs,preprintnumbers,amsmath,amssymb]{revtex4}

\usepackage{graphicx}
\usepackage{dcolumn}
\usepackage{bm}
\newcommand{\vecvar}[1]{\mbox{\boldmath$#1$}}
\newcommand{\e}{\mbox{e}}
\newcommand{\erf}{\mbox{erf}}
\newcommand{\erfc}{\mbox{erfc}}

\begin{document}

\preprint{PRESAT-7903}

\title{First-principles study on dielectric properties of NaCl crystal and ultrathin NaCl films under finite external electric field}

\author{Tomoya Ono}
\affiliation{Research Center for Ultra-Precision Science and Technology, Osaka University, Suita, Osaka 565-0871, Japan}

\author{Kikuji Hirose}
\affiliation{Department of Precision Science and Technology, Osaka University, Suita, Osaka 565-0871, Japan}

\date{\today}

\begin{abstract}
We present a first-principles study on the dielectric properties of an NaCl crystal and ultrathin NaCl films under a finite external electric field. Our results show that the high-frequency dielectric constant of the films is not affected by the finite size effect from crystal surfaces and is close to that of the crystal, whereas the static one is sensitive to the thickness of the film due to the difference in the atomic configurations between the surface and inside of the film.
\end{abstract}

\maketitle
\section{INTRODUCTION}
The response of insulators and semiconductors to external electric fields is of both fundamental and practical interest. For instance, the response of thin insulator films becomes more important as the miniaturization of microelectronics progresses. So far, alkali halide has been the subject of a large number of theoretical investigations concerning their dielectric and piezoelectric characteristics as a typical example of ionic crystals. First-principles calculations based on the density-functional theory \cite{hk,ks} have successfully been applied to studies of the linear response properties of materials at the zero bias limit \cite{linear}. However, the study of the dynamical transformations of materials in an electric field would greatly benefit from the consideration of actual {\it finite} fields. Recently, there has been a great deal of progress in the first-principles calculation for treating finite electric fields. One method is the electronic-structure calculation entirely in real-space, where the boundary conditions are not constrained to be periodic. Therefore, one can examine the properties of a material by including a finite external field potential in the Kohn-Sham Hamiltonian and imposing nonperiodic boundary conditions in the direction parallel to the field \cite{fe}. Another method is the discrete Berry phase scheme \cite{dbp1,dbp2,dbp3,dbp4,dbp5,dbp6}, in which the determinations of dielectric polarization, Born effective charges, electron-phonon interactions and several other experimentally measurable quantities of material under a finite external electric field are possible using a conventional supercell technique with a three-dimensional periodic boundary.

In this paper, we implement first-principles calculation to elucidate the dielectric properties of an NaCl crystal using the discrete Berry phase scheme and ultrathin NaCl films using the supercell with a combination of periodic and nonperiodic boundary conditions. Our results reveal that the finite size effect from crystal surfaces on the high-frequency dielectric constant of ultrathin NaCl film is negligibly small while the static dielectric constant sensitively depends on the thickness of the film because of the difference in atomic configurations between the surface and inside of the films

The rest of this paper is organized as follows. In Sec. II, we briefly describe the method used in this study. Our results are presented and discussed in Sec. III. We summarize our findings in Sec. IV. Finally, mathematical details are described in the Appendix.

\section{OVERVIEW OF COMPUTATIONAL METHOD}
Our first-principles simulations are based on the real-space finite-difference method \cite{rsfd1,rsfd2}, which enables us to determine both the self-consistent electronic ground state and the optimized geometry with a high degree of accuracy by virtue of the timesaving double-grid technique \cite{tsdg}. The norm-conserving pseudopotentials \cite{ncps} of Troullier and Martins \cite{tmpp} are used as an atomic-ion core potential. Exchange-correlation effects are treated by the local density approximation \cite{lda} in the density functional theory. The nine-point finite-difference case, i.e., the case of $N $=4 in Ref \cite{rsfd2}, is adopted for the derivative arising from the kinetic-energy operator in the Kohn-Sham equation. A cutoff energy is set at 110 Ry, which corresponds to a grid spacing of 0.30 a.u., and a further higher cutoff energy is set at 987 Ry in the vicinity of the nuclei with the augmentation of double-grid points \cite{tsdg}. The structural optimizations are carried out until the remaining forces acting on ions are smaller than 33.0 pN.

\section{RESULTS AND DISCUSSION}
\subsection{Dielectric constants of crystal}
\begin{figure}[htb]
\begin{center}
\includegraphics{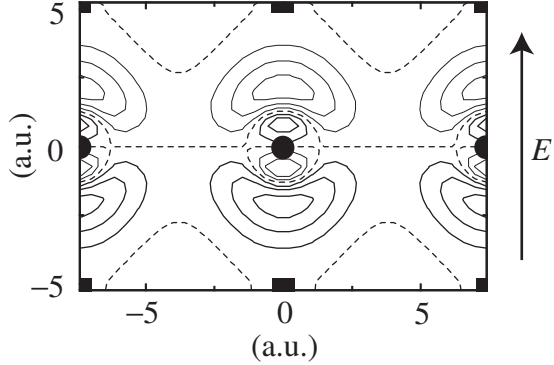}
\end{center}
\caption{Induced charge on (110) plane due to applied electric field. Contour interval is 10$^{-4}$ electrons per unit cell. Thick (Thin) curves represent positive (negative) contours and dashed curves represent zero charge contours. The atomic sites are indicated for Na by a solid square and Cl by a solid circle.}
\label{fig:bulk}
\end{figure}

The electronic band gap and dielectric constants of an NaCl crystal are calculated. The electronic ground state and the optimized geometries under external electric fields are determined using the discrete Berry phase scheme given by Umari and Pasquarello \cite{dbp5}. The calculated and experimental energy differences between the highest occupied and the lowest unoccupied levels are listed in the first column of Table~\ref{tbl:bulk}. Since the local density approximation underestimates band gaps by approximately 30 \%, there is no fundamental reason to expect a good agreement between the theoretical and experimental results. The second, third, and fourth columns of Table~\ref{tbl:bulk} show the Born effective charge, and the high-frequency and static dielectric constants of the NaCl crystal, respectively. It is well known that when an alkali halide is subjected to a static electric field, its polarization becomes larger compared with the case of the high-frequency electric field because charged ions move along the electric field lines. In theoretical calculations, if the ions are kept fixed, this yields an electric contribution under an external high-frequency electric field; if both electrons and ions are allowed to relax in response to the field, the static dielectric constant is obtained. Here, to compare the induced charge due to an applied electric field in the crystal with those in films in the next subsection, we set the external electric field to 0.25/$\varepsilon_\infty^{exp}$ V/\AA, where $\varepsilon_\infty^{exp}$(=2.25) is the high-frequency dielectric constant obtained in experiment. The electric field is applied along the [001] direction of the NaCl crystal. The calculated Born effective charge and dielectric constants are in agreement with the experimental values. Figure~\ref{fig:bulk} shows the induced charge on the (110) plane due to an applied electric field, in which ions are fixed. Since the halogen ions are polarizable units and very little charge remains on the Na sites, the primary effect of the applied electric field is the polarization of the Cl ions; the electrons are pushed from the upper side of the Cl ions to the lower side in Fig.~\ref{fig:bulk} and across the peak in the 3$p$ charge density of the Cl ions.

\begin{table}
\begin{center}
\caption{Calculated and experimental values of the electronic band gap $E_{gap}$, Born effective charge $Z^*$, high-frequency dielectric constant $\varepsilon_\infty$ and static dielectric constant $\varepsilon_0$ of NaCl crystal.}
\label{tbl:bulk}
\begin{tabular}{c|cccc} \hline\hline
& \hspace {3mm} $E_{gap}$ (eV) \hspace {3mm} & \hspace {3mm} $Z^*$ \hspace {3mm} & \hspace {3mm} $\varepsilon_{\infty}$ \hspace {3mm} & \hspace {3mm} $\varepsilon_0$ \hspace {3mm} \\ \hline
This work  & 4.55 & 1.11 & 2.31 &  5.70 \\
Experiment    & 8.6\tablenotemark[1]  & 1.12\tablenotemark[2] & 2.25\tablenotemark[3] &  5.62\tablenotemark[3]  \\ \hline \hline
\end{tabular}
\tablenotetext[1]{From Ref.\cite{nakai}.}
\tablenotetext[2]{From Ref.\cite{mei}.}
\tablenotetext[3]{From Ref.\cite{dekker}.}
\end{center}
\end{table}

\subsection{Dielectric constants of films}
\begin{figure}[htb]
\begin{center}
\includegraphics{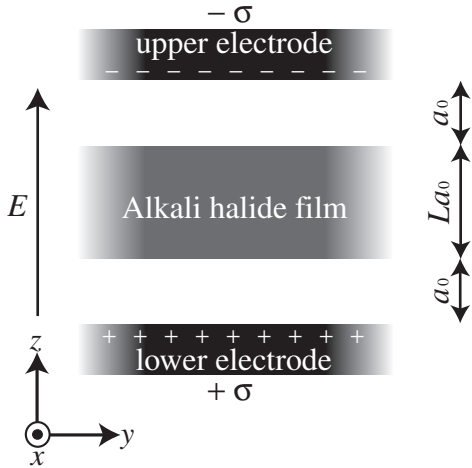}
\end{center}
\caption{Computational model for thin films. The film is sandwiched between charged electrodes.}
\label{fig:model}
\end{figure}

\begin{figure}[htb]
\begin{center}
\includegraphics{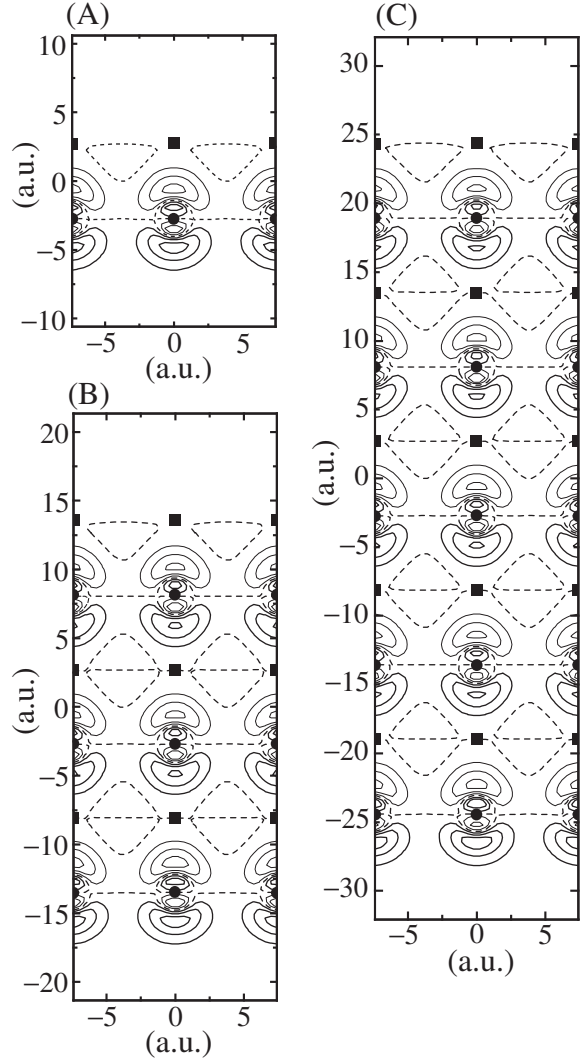}
\end{center}
\caption{The induced charge on (110) plane due to the applied electric field for (A) $L=1$, (B) $L=3$, and (C) $L=5$. Contour interval is 10$^{-4}$ electrons per unit cell. The meanings of the symbols are the same as those in Fig. \ref{fig:bulk}.}
\label{fig:film}
\end{figure}

The dielectric constants of ultrathin NaCl films with (001) surfaces are examined. In the calculations for the films, the model shown in Fig.~\ref{fig:model} is employed. The supercells are taken to be $ a_0 \times a_0 \times (L+2)a_0 $ and periodic boundary conditions are imposed in the $x$- and $y$-directions and a nonperiodic boundary condition is imposed in the $z$-direction, where $a_0$(=10.66 a.u.) is the lattice constant of an NaCl crystal obtained in experiments and  $L$ is the number of lattices in the films. The ionic potential and the ionic Coulomb energy under the two-dimensional periodic boundary condition are described in the Appendix. The Coulomb charge density $\sigma$($=3.87 \times 10^{-4}$ $e$/bohr$^2$) on the upper and lower electrodes is determined so that the electric field, $E$, is 0.25 V/\AA \hspace{2mm} when there is no material between the electrodes. We first examine the optimized atomic geometries of the NaCl films in the absence of the external electric field: The Na atom layers on both the surfaces are found to sink toward the inside of the films and locate inner than Cl atom layers by $\sim$ 0.1 a.u. Subsequently, the atomic and electronic structures of the films under the external electric field are explored. Figure~\ref{fig:film} shows the induced charges on the (110) plane in the cases of $L=1$, 3 and 5, in which ions are kept frozen. There are no marked differences in the polarizations around the Cl ions between the thick films and thin films as well as between the films and the crystal.

\begin{table}
\begin{center}
\caption{Calculated values of electronic band gap and dielectric constants.}
\label{tbl:film}
\begin{tabular}{c|ccc} \hline\hline
& \hspace {5mm} $E_{gap}$ (eV) \hspace {5mm} & \hspace {5mm} $\varepsilon_{\infty}$ \hspace {5mm} & \hspace {5mm} $\varepsilon_0$ \hspace {5mm} \\ \hline
$L=1$ & 4.81 & 1.95 & 2.75 \\
$L=2$ & 4.69 & 2.07 & 3.24 \\
$L=3$ & 4.63 & 2.15 & 3.62 \\
$L=4$ & 4.58 & 2.21 & 3.99 \\
$L=5$ & 4.55 & 2.26 & 4.36 \\ \hline \hline
\end{tabular}
\end{center}
\end{table}

\begin{figure*}[htb]
\begin{center}
\includegraphics{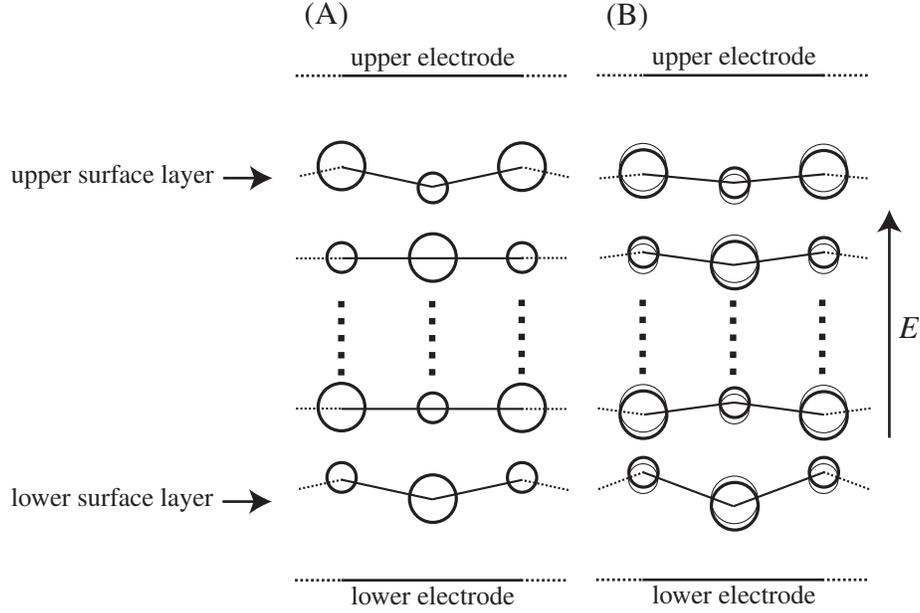}
\end{center}
\caption{Schematic views of deformations of atomic layers on (010) plane of films for the cases of (A) the absence of an external electric field and (B) its presence. Large and small circles are Cl and Na ions, respectively.}
\label{fig:deformation}
\end{figure*}

The computed band gap, high-frequency dielectric constant and static dielectric constant of the NaCl films are listed in Table~\ref{tbl:film}. The dielectric constants are evaluated according to
\begin{equation}
\varepsilon=\frac{-P}{4 \pi E La_0^3} +1 ,
\end{equation}
where $P$ and $E$ are the dielectric polarization and the average electric field between the electrodes in the $z$-direction, respectively \cite{comment1}. The band gap of the film becomes smaller and approaches that of the crystal, as the thickness of the film increases. The high-frequency dielectric constant of the thin films does not significantly vary with the thickness of the films and is close to that of the crystal, which account for the results shown in Fig.~\ref{fig:bulk} and Fig.~\ref{fig:film} where there is no notable difference in the induced charge around the Cl ions even though the ions locate on the surface of the films. These results represent that the high-frequency dielectric constant is dominated only by the polarization of the Cl ions. On the other hand, the static dielectric constant of the thin films markedly changes according to the film thickness. This is interpreted as being due to the difference in atomic configurations between the surface and inside of the film; on the surface atomic layer of the upper side, Cl ions are closer to the upper electrode than the Na ions, while on the other atomic layers, the Na ions are closer to the upper electrode than the Cl ions (see Fig.~\ref{fig:deformation}). These results indicate that the atomic configurations of the films under the finite external electric field play important roles in the response of the films to the static electric field.

\section{CONCLUSION}
We have studied the dielectric properties of the NaCl crystal and ultrathin NaCl films under a finite external electric field. When the external electric field is applied, electrons are pushed across the peak in the 3$p$ charge density of the Cl ions in both the crystal and the films. The high-frequency dielectric constant is hardly affected by the finite size effects due to the crystal surfaces and is close to that of the crystals. This is consistent with the induced charge due to the electric field, in which no significant difference in the induced charge around the Cl ions is observed between the thick films and thin films as well as between the films and the crystal. On the other hand, the static dielectric constant is sensitive to the thickness of the film because of the difference in the atomic configurations between the surface and inside of the film. The other alkali-halide materials are expected to emerge the similar properties to that of NaCl and the work in progress is the investigation for them.

\section*{ACKNOWLEDGMENTS}
This research was supported by a Grant-in-Aid for the 21st Century COE ``Center for Atomistic Fabrication Technology'' and also by a Grant-in-Aid for Scientific Research (C) (Grant No. 16605006) from the Ministry of Education, Culture, Sports, Science and Technology. The numerical calculation was carried out by the computer facilities at the Institute for Solid State Physics at the University of Tokyo, and the Information Synergy Center at Tohoku University.

\section*{APPENDIX: COULOMB POTENTIAL AND ENERGY UNDER 2D PERIODIC AND 1D NONPERIODIC BOUNDARY CONDITIONS}
The ionic potential and ionic Coulomb energy in the cases of two-dimensional periodic and one-dimensional nonperiodic boundary conditions are obtained using the Ewald summation. Here, the case in which periodic boundary conditions are imposed in the $x$- and $y$-directions and an isolated boundary condition is imposed in the $z$-direction is explained.

By using the identity ($a, b, \eta>0$)
\begin{eqnarray}
\lefteqn{\int^\eta_0 t^{-2} \e^{-a^2t^2-\frac{b^2}{4t^2}}dt} \nonumber \\
&\hspace{0.3cm}=&\frac{\sqrt{\pi}}{2b}\left[\e^{-ab}\erfc\left(\frac{b-2\eta^2a}{2\eta}\right)+\e^{ab}\erfc\left(\frac{b+2\eta^2a}{2\eta}\right)\right], \nonumber \\
\end{eqnarray}
the ionic pseudopotential is given by
\begin{widetext}
\begin{eqnarray}
\label{eqn:ono-ewald2D01}
V^s_{loc}(\vecvar{r})&=&-\frac{\pi}{S} Z_s \sum_{\left|\vecvar{G}\right| \ne 0} \frac{1}{|\vecvar{G}|} \cos \left[ \vecvar{G} \cdot (\vecvar{r} - \vecvar{R}^s) \right] f^+(\vecvar{G},\vecvar{r}) \nonumber \\
&&+\frac{2\sqrt{\pi}}{S} Z_s \left[ \frac{1}{\eta} \e^{-|z-R_z^s|^2\eta^2}+\sqrt{\pi} \: |z-R_z^s| \, \erf (|z-R_z^s| \: \eta) \right] \nonumber \\
&&+Z_s \sum_{\vecvar{P}} \frac{1}{|\vecvar{\zeta}^s|} \Biggl[ \erf (\eta \: |\vecvar{\zeta}^s|)-\sum_{i=1,2} C_{s,i} \, \erf (\sqrt{\alpha_{s,i}} \: |\vecvar{\zeta}^s|) \Biggr],
\end{eqnarray}
\end{widetext}
where $\vecvar{G}=2\pi(\frac{j_x}{L_x},\frac{j_y}{L_y},0)$, $\vecvar{P}=(n_xL_x,n_yL_y,0)$, $S=L_x \times L_y$, $\vecvar{\zeta}^s=\vecvar{P}+\vecvar{r}-\vecvar{R}^s$, $\vecvar{R}^s$ is the position of the $s$-th ion, $L_x$, $L_y$, and $L_z$ are the lengths of the unit cell in the $x$-, $y$-, and $z$-directions, respectively, and
\begin{eqnarray}
\label{eqn:ono-ewald2D02}
f^+(\vecvar{G},\vecvar{r})&=& \e^{-\left|\vecvar{G}\right| \: (z-R_z^s)} \erfc \left( \frac{\left|\vecvar{G}\right|-2\eta^2(z-R_z^s)}{2\eta} \right) \nonumber \\
&&+ \e^{\left|\vecvar{G}\right| \: (z-R_z^s)} \erfc \left( \frac{\left|\vecvar{G}\right|+2\eta^2(z-R_z^s)}{2\eta} \right). \nonumber \\
\end{eqnarray}
In addition, $C_{s,i}$ ($C_{s,1}+C_{s,2}=1$) and $\alpha_{s,i}$ are the parameters of the pseudopotential given by Hamann {\it et al}. \cite{hsc}, the sum of $\sum_{\vecvar{G} \ne 0}$ is taken over the reciprocal lattice vectors excluding the case of $|\vecvar{G}|=0$, $\erf(x)$ is the error function (or probability integral) defined by $\erf(x)=\frac{2}{\sqrt{\pi}}\int^x_0 \e^{-t^2}dt$, and $\erfc(x)=1-\erf(x)$. The summations of the first and third terms in Eq.~(\ref{eqn:ono-ewald2D01}) are taken over the reciprocal lattice vectors $\vecvar{G}$ and the real-space lattice vectors  $\vecvar{P}$.  When $\eta$ is chosen to be 0.2 -- 0.7, the number of iterations in Eq. (\ref{eqn:ono-ewald2D01}) is only $7^2$ -- $11^2$.

Then, the ionic Coulomb energy is given by
\begin{widetext}
\begin{eqnarray}
\label{eqn:ono-ewald2D14}
\gamma_E &=& \frac{\pi}{2S} \sum_{s,s'} Z_s Z_{s'} \sum_{\left|\vecvar{G}\right| \ne 0} \frac{1}{|\vecvar{G}|} \cos \left[ \vecvar{G} \cdot (\vecvar{R}^{s'} - \vecvar{R}^s) \right] f^+(\vecvar{G},\vecvar{R}^{s'}) - \sum_{s,s'} Z_s Z_{s'} \delta_{ss'} \frac{\eta}{\sqrt{\pi}} \nonumber \\
&&-\frac{\sqrt{\pi}}{S} \sum_{s,s'} Z_s Z_{s'} \left[ \frac{1}{\eta} \e^{-|R_z^{s'}-R_z^s|^2 \eta^2} + \sqrt{\pi} \: |R_z^{s'}-R_z^s| \, \erf (|R_z^{s'}-R_z^s| \eta) \right] \nonumber \\
&&+\frac{1}{2} \sum_{\vecvar{P},s,s'} \!\!\!\mbox{}^\prime Z_s Z_{s'} \frac{\erfc (\eta \: |\vecvar{\xi}^{s,s'}|)}{|\vecvar{\xi}^{s,s'}|} ,
\end{eqnarray}
\end{widetext}
where $\vecvar{\xi}^{s,s'}=\vecvar{P}+\vecvar{R}^{s'}-\vecvar{R}^s$. A nucleus does not interact with its own Coulomb charge, so the $\vecvar{P}=0$ term can be omitted from the real-space summation when $s=s'$. The prime in the last summation means that $|\vecvar{\xi}^{s,s'}|=0$ is omitted.


\begin{references}
\bibitem{comment1}
Since there is no definite way of determining the thicknesses of the films, here we assume them to be $La_0$.
\end{references}
\end{document}